\documentstyle[preprint,prb,epsf,aps]{revtex}
\begin{document}
\draft

\title{Calculation of Giant Magnetoresistance in Laterally
 Confined Multilayers}
\author{Kingshuk Majumdar, Jian Chen, and Selman Hershfield}
\address{Department of Physics and National High Magnetic
Field Laboratory, University of Florida,
215 Williamson Hall, Gainesville, FL 32611}
\date{\today}
\maketitle

\begin{abstract}
We have studied the Giant Magnetoresistance (GMR) for laterally confined
multilayers, e.g., layers of wires, using the classical Boltzmann 
equation in the current-in-plane (CIP) geometry.  
For spin-independent specularity factors at the sides of the wires
we find that the GMR due to bulk and surface scattering decreases with 
lateral confinement. The length scale at which this occurs is of order the
film thickness and the mean free paths. 
The precise prefactor
depends on the relative importance of surface and bulk scattering anisotropies.
For spin-dependent specularity factors at the sides of the wires
the GMR can increase in some cases with decreasing width.  
The origin of the change in the GMR in both cases can be understood in
terms of lateral confinement changing the effective mean free paths 
within the layers.
\end{abstract}

\bigskip

\section{INTRODUCTION}
	
	Electrical transport properties of magnetic multilayers, which
are thin alternating layers of ferromagnets (FM) and paramagnets (PM), has 
drawn considerable interest in recent years.~\cite{ptoday}  
A large decrease in the resistivity from antiparallel to 
parallel alignment of the film magnetizations has been observed 
experimentally.~\cite{{baibich},{grunberg}}
This phenomenon is known as the giant magnetoresistance (GMR). 
The decrease in the resistivity from antiparallel to parallel alignment
arises from spin-dependent scattering.~\cite{baibich} 
The two sources of spin dependent scattering in these
multilayers are bulk scattering and surface scattering.  
There have been numerous experimental and theoretical studies
to understand the physics of the giant magnetoresistance
and to use it in applications.\cite{{ptoday},{white}}
 
 There are several theoretical approaches to transport in magnetic multilayers.
One approach is to use the phenomenological Fuchs-Sondheimer 
theory of thin film resistance.~\cite{{fuchs},{sond}} This approach is 
based on the classical Boltzmann equation. Other approaches are based on 
the linear response theory~\cite{{levy},{levy2},{fert}} and the 
quantum Boltzmann equation.\cite{visscher} It has been shown 
that the conductivities obtained from the classical Fuchs-Sondheimer theory
are in good agreement with the quantum results obtained via the Kubo 
formula.\cite{zhang} Thus it is useful to use the semiclassical approach 
to understand the physics of the giant magnetoresistance. 
Furthermore, this method with spin-dependent interface scattering\cite{c-b}   
reproduces the qualitative features of the giant magnetoresistance seen
in the experiments.~\cite{{baibich},{grunberg}}
   
With the development of nanotechnology, it is becoming
important to understand the effect of lateral confinement on the 
GMR.~\cite{childress}  In this paper, we address the following 
questions: (i) Does the GMR increase or decrease with the reduction in
width? (ii) If there is an increase or decrease in the GMR what causes
it? (iii) What are the relevant length scales in the problem?
We use the classical Boltzmann equation to answer the above questions. 
   
     	The layout of the rest of our paper is as follows. 
  In Sec. II a detailed 
  description of the model and our numerical procedure is given. 
  In Sec. III we present the results of our calculation, and  
  Section IV contains the conclusion.
  
 \section{THE MODEL}
 
 The geometry of our problem is shown in Figure 1.
 We compute the current-in-plane conductivity (CIP) for a three-wire
 structure stacked along the $\hat y$-direction. 
 Both the current and the 
 electric field are in the $\hat z$ direction. We have two identical
 ferromagnetic materials (FM) and a paramagnet (PM).
 For this three wire geometry we compute the conductivity in the case 
 when the ferromagnet's magnetizations are parallel, 
 $\sigma _{F}$, and when they are
 antiparallel, $\sigma _{AF}$.  
  The magnetoresistance is defined as the ratio of the change in the 
  conductivity from parallel to antiparallel alignments 
  divided by the parallel conductivity:
\begin{equation}
	{\rm GMR} = 1- \frac{\sigma_{AF}}{\sigma _{F}}. 
	\label{GMR}
\end{equation}

  		To calculate the conductivities for different spin alignments
 we consider the classical steady state Boltzmann transport equation: 
 \begin{equation}
	{\bf v \cdot \nabla_r}f({\bf v, r})
	-\frac {e}{m}
		{\bf E \cdot \nabla_v} f({\bf v, r})
	=\left(\frac {\partial f({\bf v, r})}{\partial t}\right )_{scatt},
\label{boltz}
\end{equation}
where $f({\bf v, r})$ is the distribution function for electrons of mass
$m$ at position ${\bf r}$ with velocity ${\bf v}$ in presence of the
 electric field ${\bf E}$. Because of the
 scattering term, Eq. (\ref{boltz}) is complicated. For simplicity 
 we consider the case where the scattering term is:
 \begin{equation}
  	\left(\frac {\partial f({\bf v, r})}{\partial t}\right)_{scatt} =
  	 -\frac {f({\bf v, r})-\left < f({\bf v, r})\right >}{\tau},
 \label{approx}
 \end{equation}
 with
 $\left<f({\bf v, r})\right>$ being the spherical average of the 
 distribution function,
 and $\tau$ is the relaxation time.  This is the simplest scattering
 term to represent elastic scattering.
 To solve Eq. (\ref{boltz}) with the right hand side of Eq. (\ref{approx}),
 we define $g({\bf v}, {\bf r})$ to be the deviation of the 
 distribution function from its equilibrium value:
 \begin{equation}
 	f({\bf v, r}) = f^{eq}({|\bf v}|)+g({\bf v}, {\bf r}).
 \label{dist-func}
 \end{equation}
 We next make the ansatz
 that within linear response the spherical average of the distribution
 function is equal to the equilibrium distribution function,
 $\left<f({\bf v, r})\right>=f^{eq}(|{\bf v}|)$, i.e.,
 the spherical average of $g({\bf v}, {\bf r})$ is zero.
 This will be checked explicitly later.
 The linearized Boltzmann equation for a wire labeled by an index $n$ is then
 given by
\begin{equation}
{\bf v \cdot \nabla_r} g_{n s}({\bf v,  r})
 - \frac e{m}{\bf E \cdot \nabla_v} f^{eq}_{n s}(|{\bf v}|)
  = -\frac {g_{n s}({\bf v, r)}}{\tau_{n s}},
\label{diffeqn}
\end{equation}
where $s$ denotes the spin of the electrons.
When the electric field is zero, Eq. (\ref{diffeqn}) 
becomes a homogeneous
equation with the expected solution being $g_{ns}({\bf v, r})=0$.  For
non-zero electric field, $g_{ns}({\bf v, r})$ is proportional to 
$(e {\bf E}/m) {\bf \nabla_v}f_{ns}^{eq}(|{\bf v}|)
=e {\bf E \cdot v} \partial f_{ns}^{eq}(|{\bf v}|) / \partial \epsilon_{v}$.
Thus, $g_{ns}(-{\bf v, r})=-g_{ns}({\bf v, r})$, irrespective  
of the boundary conditions, and the ansatz 
that the spherical average of $g_{ns}({\bf v, r})$ is zero is justified.

The general solution of the above equation\cite{chambers} is
\begin{eqnarray}
	g_{n s}({\bf v, r}) &=& g_{n s}
	({\bf v, r_B})e^{-|{\bf r -r_B}|/\tau_{n s}|{\bf v}|} 
	\nonumber \\
	&+& e\tau_{n s}{\bf E \cdot v}\left(\frac{\partial f^{eq}_{n s}}
		{\partial \epsilon_v}\right)(1-e^{-|{\bf r -r_B}|/
		\tau_{n s} |{\bf v}|}),\nonumber \\
	& &
\label{solution}
\end{eqnarray}
where 
 ${\bf r}_B$ is a point on the boundary or interface.
  Eq. (\ref{solution}) implies that to find the distribution
 function at position ${\bf r}$ with velocity ${\bf v}$, 
 we proceed from ${\bf r}$
 backwards along ${\bf v}$ until we reach a point ${\bf r}_{B}$ at the 
 boundary. The distribution function at
 the boundary, $g_{n s}({\bf v, r_B})$, is 
 determined by the boundary conditions. We recover the usual bulk
 value if we go far away from the boundary points.
 Also we can see from Eq. (\ref{solution}) that
 the electrons lose their momentum as they diffuse into the medium and the 
 characteristic length scale for this is just the mean free path,
 $\tau_{ns} v_F$, where $v_F$ is the Fermi velocity. 
 
 	We now examine the boundary conditions. At each interface the 
 electrons undergo either specular or diffuse scattering. 
 For the n\underline{th} interface, which is between wires $n$ and $n+1$,
 we define the probability of spin $s$ electrons
 being diffusively scattered as $(1-S_{ns})$, where $S_{ns}$ is the 
 spin dependent specularity factor.
 The probabilities for being specularly reflected and transmitted are
 $S_{ns}R_{ns}$ and $S_{ns}T_{ns}$, respectively. 
 The sum of $R_{ns}$ and $T_{ns}$ is one.
 The angular dependence of the surface scattering parameter,~\cite{lenk} the 
  reflection coefficients,
  and the transmission coefficients~\cite{{hood},{stiles}} has been
 studied, but in our calculation we treat those as angle
  independent.  With these definitions, 
  the boundary conditions at the n{\underline {th}} interface 
  can be expressed as: 
\begin{eqnarray}
 g^{out}_{n s}({\bf v},x,y_n) & = & S_{n s}T_{n s}g^{in}_{n-1 s}
 				({\bf v},x,y_n) \nonumber \\
 			&+& S_{n s}R_{n s}g^{in}_{n s}
 				({-\bf v},x,y_n),
 \label{b1} \\
 g^{out}_{n-1 s}({-\bf v},x,y_n) & = & S_{n s}T_{n s}g^{in}_{n s}
 				({-\bf v},x,y_n)  \nonumber \\
 			&+& S_{n s}R_{n s}g^{in}_{n-1 s}
 				({\bf v},x,y_n),
 \label{b2}
\end{eqnarray}
where $y_n$ is the position of the interface, and the superscripts 
$out, in$ correspond to electrons going out from the
boundary or coming in to the boundary.
Similar equations are satisfied at the sides of the wires except
that there is no transmission.

 We use an iterative procedure to compute the distribution 
functions, $g_{n s}^{out}({\bf v, r_B})$, which are non-uniform along the 
$\hat x$ and $\hat y$ directions of
the interfaces and edges (see Fig. 1).  
From Eq. (\ref{diffeqn}), we observe that the
$g_{n s}^{out}({\bf v, r})$'s with different velocities projected along
the $\hat z$ direction are decoupled.  This allows us to discretize the
$g_{n s}^{out}({\bf v, r})$ according to $\cos \theta = v_{z}/|v|$ and
solve each separately.  For each $\cos \theta$, $g_{n s}^{out}({\bf v, r})$
at the edges and the interfaces carry two more indices: an angle
$\phi$ and a position ${\bf r}_i$. The 
$g_{n s}^{out}({\bf v, r})$'s for different 
$\phi$ and ${\bf r}_i$ are related via Eq. (\ref{solution}) and the boundary
conditions given in Eqs. (\ref{b1}) and (\ref{b2}). To illustrate this
 we consider the relations between the distribution functions
 of the outgoing electrons at the first interface in Fig. 1:
\begin{eqnarray}
& &g^{out}_{1 s}(\cos \theta, \phi, {\bf r}_j) = S_{1 s} R_{1 s}
 g^{out}_{1 s}(\cos \theta, 2\pi - \phi, {\bf r}_i) \nonumber \\
&\times & e^{-d_1/\tau_{1 s}v_F \sin \theta} 
+ S_{1 s}T_{1 s} g^{out}_{2 s}(\cos \theta, \phi, {\bf r}_k) e^{-d_2/
\tau_{2 s} v_F \sin \theta}, \nonumber \\
& &
\label{numerics}
\end{eqnarray}
where $d_1, d_2$ are the path lengths projected onto the $x-y$ plane
 from one boundary to another. 
Initial guesses for $g_{n s}^{out}
(\cos \theta, \phi, {\bf r}_i)$ are taken from the nearby 
$g_{n s}^{out}(\cos \theta - \delta (\cos \theta),\phi, {\bf r}_i)$.
The calculations are converged to within 1\% with the total 
number of divisions for
$\cos \theta$, ${\bf r}_i$, and $\phi$ chosen as $N_{\cos \theta}$=100,
$N_{\phi}$=500, and $N_{i}$=400, respectively.

	Once the distribution functions at the boundaries are known, by
 using Eqs. (\ref{dist-func}) and (\ref{solution})  the distribution 
 function of the electrons with momentum ${\bf v}$ at any point ${\bf r}$
 can be determined. We explicitly check that our ansatz is valid:
 $\langle g_{ns}({\bf v, r})\rangle =0$.
 The current density along the direction of the electric field for wire
 $n$ and spin $s$  is:
 \begin{equation}
 {\cal J}_{ n s}({\bf r}) = -e\left(\frac{m}{h} \right)^3 
 \int\! v_z g_{n s}({\bf v, r}) d^3 {\bf v}.
 \end{equation}
 The conductivity is obtained by averaging the current density over
 a cross-sectional area, ${\cal A}$:
 \begin{equation}
 \sigma = \frac1{E{\cal A}}\sum_{n=1}^{3} \sum_{s = \uparrow, \downarrow}
 \int \! {\cal J}_{n s} (x,y) dx dy .
 \label{sigma}
 \end{equation}
 Finally, the
 giant magnetoresistance is obtained from Eq. (\ref{GMR})
 using parallel and antiparallel alignments of the magnetizations
 in Eq. (\ref{sigma}).

\section {NUMERICAL RESULTS}

	We begin this section with the different input parameters of
 our problem. 
 They are: 
 the spin dependent mean free paths in the ferromagnet, 
  $L^{\uparrow ,\downarrow}$,
 the mean free path in the paramagnet, $L^{\rm PM}$,
 the spin-dependent transmission coefficients, $T^{\uparrow ,\downarrow}$, 
 the specularity factors, $S$, 
 and the thicknesses of the layers,
 $t_{\rm FM}$ and $t_{\rm PM}$.
 We choose the mean free paths to be those determined in an experiment
 on a Co/Cu/Co structure:\cite{gurney} $L^\uparrow$=55\AA, 
 $L^\downarrow$= 10\AA, and $L^{\rm PM}$= 226\AA.
 The transmission coefficients are obtained from an average of the
 transmission coefficients calculated by
 Stiles:\cite{stiles} $T^{\uparrow} = 0.8$ and $T^{\downarrow} = 0.4$.
 We take the specularity factors at all interfaces and sides
 to be the same: $S =0.9$.  Finally, the thicknesses are chosen to be
 those of Cu when the Co layers are antiferromagnetically coupled at zero 
 field.\cite{parkin}  For simplicity we take  the ferromagnets to have
 the same thicknesses as the paramagnets: $t_{\rm FM} = t_{\rm PM}$.

 With the above parameters the giant magnetoresistance is due to both
 spin anisotropies in the bulk and surface scattering.  It is useful
 to consider the limiting cases when the GMR is due to only bulk scattering
 anisotropies or only surface scattering anisotropies.  To do this,
 we consider two special cases: (a) when the transmission coefficients
 are equal: $T^{\uparrow} = T^{\downarrow} = 0.6$ and (b) when the
 bulk mean free paths in the ferromagnet are equal: 
 $L^{\uparrow} = L^{\downarrow} = 17$\AA.  In the following these are 
 referred 
 to as the (a) bulk scattering case and (b) surface scattering case.
 
 In Fig. 2 we plot the giant magnetoresistance as a function of width
 for the three cases:
 (a) GMR due to bulk scattering anisotropies, 
 (b) GMR due to surface scattering anisotropies, and 
 (c) GMR due to both bulk and surface scattering anisotropies.
 In all cases the giant magnetoresistance
 decreases as we reduce the width. 
 To understand this decrease with lateral confinement we consider
 a slab, i.e., a  wire with infinite width.
 Diffusive scattering 
 at the sides of the wire reduces  
 the conductivity and mean free paths of the wire compared to the slab.
 This reduction
 in the conductivity in going from a slab to a wire should be comparable
 to the reduction in going from a bulk system to a slab with a thickness 
 equal to $w$. The conductivity of such a slab is 
 given by\cite{sond}
 \begin{eqnarray}
 \sigma_{\rm slab} &=& \left(\frac{ne^2L}{m v_F}\right)\Bigg\{1-\frac{3L}
 {2w}(1-S)\nonumber \\
 	&\times &
 	\int_0^1 \!d(\cos \theta) \cos \theta \sin^2 \theta
 	\frac{1-\exp(-w/L|\cos\theta|)}
 	{[1-S\exp(-w/L |\cos\theta|)]}\Bigg\},
 \label{slab}
 \end{eqnarray}
 and the bulk conductivity is
 \begin{equation}
 \sigma_{\rm bulk} = \left(\frac{ne^2L}{m v_F}\right),
 \label{bulk}
 \end{equation}
 where $L$ is the mean free path of the electrons.
 We obtain an
 effective mean free path, $L_{\rm eff}$, 
 by replacing $L$ in Eq. (\ref{bulk}) with $L_{\rm eff}$
 such that $\sigma_{\rm bulk}=\sigma_{\rm slab}$.
 The effective mean free paths can be used in a
 multilayer calculation with infinite width wires.
 The results of such an effective mean free path multilayer calculation
 are plotted as the solid lines in Fig. 2.  
 This approximation is in good agreement with the exact results (symbols),
 and hence we conclude that the reduction in the GMR is due to 
 a decrease in the effective mean free path within the layers.

 Although the GMR is reduced in all the cases shown in Fig. 2,
 the actual GMR vs. width curves are different.  In particular,
 there are different length scales at which the GMR is reduced.
 The width at which the GMR is half its infinite width value,
 GMR(wire)/GMR(slab) $=1/2$, is defined as the half-width.
 In Fig. 3 we have rescaled the GMR vs. width curves shown
 in Fig. 2 by GMR(slab) and the half-width.  All the points
 fall close to a single curve.  This means that the half-width
 and the GMR for infinite width wires, GMR(slab), determine the
 GMR vs. width curves.  We have also tried the same rescaling with
 other values of $S$, which is still the same for all sides
 and interfaces, and found the same curve.

 In Figs. 4(a) - 4(c) we have plotted 
 the half-width as a function of the mean free paths 
 in the ferromagnet and the thicknesses. 
 The cases (a) - (c) refer to the same parameters as in Fig. 2.
 The ratio between the mean free paths in the ferromagnet is fixed to
 $L^{\uparrow}/L^{\downarrow}$=5.5 in Figs. 4(a) and (c), 
 which is the same ratio used in Figs. 2(a) and (c).
 As for Fig. 4(b), the mean free paths are equal 
 in the surface scattering case of Fig. 2(b):
 $L^{\uparrow} = L^{\downarrow}$.
 For bulk scattering anisotropies (Fig. 4(a)) the half-width increases almost
 linearly with both the thickness and the mean free path.  
 On the other hand, for surface scattering anisotropies (Fig. 4(b))
 the half-width depends primarily on the thickness of the film
 and only weakly on the mean free paths in the ferromagnet.
 When both surface and bulk scattering anisotropies are present (Fig. 4(c)),
 the dependence of the half-width on the mean free path and thickness
 can be complicated.  In the region where the half-width has a peak in
 Fig. 4(c), the GMR has a local minimum as a function of the mean free path.  
 The origin of this local minimum comes from the near cancelation of the GMR 
 from the bulk and surface contributions.  The location of this 
 minimum depends primarily on the transmission coefficients.

 Although we believe the generic behavior is that the GMR will
 decrease when the multilayers are laterally confined, one can find
 parameters where the GMR actually increases with lateral confinement.
 The two ways we have found to do this are (i) have the sides of the wires
 introduce additional spin dependence in the scattering and (ii)
 have the sides of the wires selectively decrease the resistivity 
 in the ferromagnet relative to the paramagnet.  The idea behind
 both of these is again that laterally confining the multilayers
 reduces the effective mean free paths within each layer.
 By changing the mean free paths by different amounts, one can
 tune the GMR.
 As an example in Fig. 5 we have plotted the GMR vs. width for
 two cases which differ only by the specularity factor at the sides
 of the FM.  For the solid curve the specularity factors are 
 0.9 for both spin-up and spin-down electrons, while for the dashed curve
 the specularity factor at the sides for spin-up electrons is 0.9 and for
  spin-down electrons
 is 0.5.  The GMR actually increases as one decreases the width
 of the sample because the mean free path for spin-down electrons
 decreases more rapidly than the mean free path for spin-up 
 electrons.  Such spin dependent specularity factors may occur naturally
 or be attainable by coating the sides of the multilayer.
 Allowing spin-dependent specularity factors at non-transmitting 
 interfaces opens
 up the possibility that one can create a GMR device without 
 ferromagnetic conductors, but only insulating ferromagnets which
 change the surface scattering.
 
\section {CONCLUSION}

 	In this paper we have studied the effect of lateral
 confinement on the giant magnetoresistance. 
 For spin independent specularity factors at the sides
 of the wires, the GMR decreases as one reduces the width of the wires.
 For this case we found that the GMR vs. width curves could be 
 very nearly collapsed
 onto a single curve by rescaling the GMR by its infinite width value
 and the width by the half-width.  The half-width depends on both the
 mean free paths and the thickness of the films.  In the case of the
 GMR due solely to surface scattering, the thickness of the sample
 plays a dominant role in determining the half-width, while in the
 case of GMR due solely to bulk scattering, the half-width increases
 roughly linearly with both the mean free paths in the ferromagnetic wires
 and the thicknesses.  The general case when both surface and bulk 
 anisotropies are important can lead to more complex dependencies.

 The source of the
 decrease in the GMR is a reduction in the effective mean free
 path in the layers due to scattering off the sides of the wires.
 We showed this quantitatively by determining an effective mean free
 path within each wire and substituting it into a multilayer 
 calculation for films (infinite width wires).  The results of the 
 approximate solution and the exact solution agree quite well.
 
 For the case of spin-dependent mean free paths we find that there
 are parameter regimes where the GMR can increase as the sample
 is laterally confined.  The origin of the increase is again changing
 of the effective mean free path within each layer as one decreases
 the width.  With spin-dependent specularity factors one can 
 change the ratio of the mean free paths for spin-up and spin-down
 electrons and hence change the GMR.  Thus, with appropriately
 prepared sides of the wires, one may be able to increase or at least
 stem the decrease in the GMR.
 In any case the length scale at which the reduction in the GMR
 takes place is typically quite small, of order the mean free paths
 and the thicknesses of the layers. The effects discussed here
 will not be important until one goes to very small samples.  
 
 	This work was supported by DOD/AFOSR grant F49620-96-1-0026 and
 NSF frant DMR9357474, and the NHMFL. We thank J. Childress, T. S. Choy
 for helping us in different stages of our work.

\begin{figure}
\protect \centerline{\epsfxsize=3.0in \epsfbox {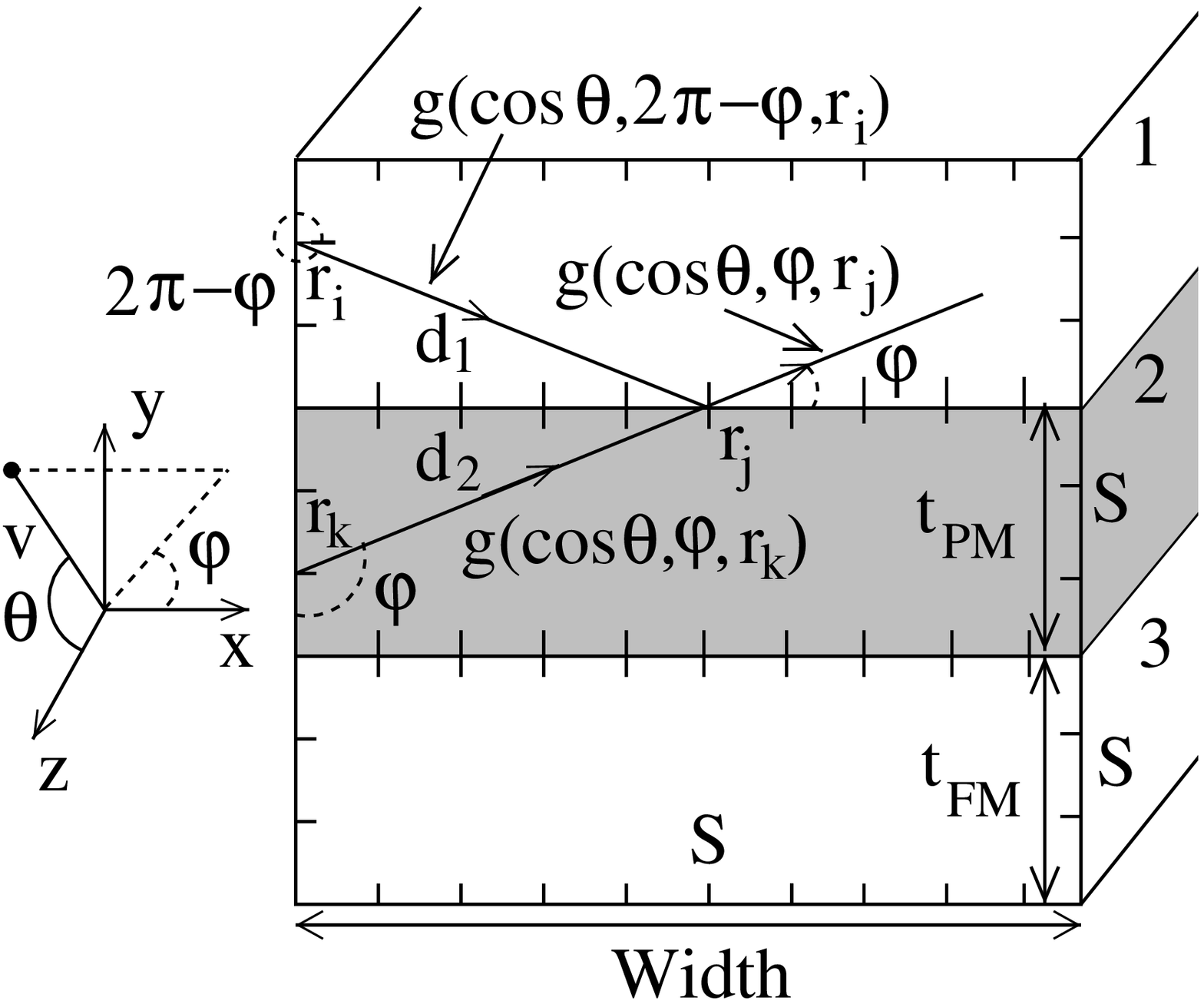}}
\vskip .6cm
\protect\caption{
 Schematic diagram of the three wire structure. Wires 1 and 3 are the 
 ferromagnets of thickness $t_{\rm FM}$, and wire 2 is a paramagnet of
 thickness $t_{\rm PM}$. Current is in 
 plane (CIP) along the ${\hat z}$ direction, and the wires are stacked in
 the $\hat y$ direction. $S$ represents the spin independent
 specularity factors at the sides of the ferromagnet,
 and the paramagnet.
 To determine the distribution function for outgoing electrons 
 at point ${\bf r}_j$, $g(\cos \theta ,\phi ,{\bf r}_j)$,
 one must consider all possible incoming
 electrons from points on the edges and interfaces, e.g., points
 ${\bf r}_i$ and ${\bf r}_k$.  The precise relationship is described in the
 text (see Eq. (\ref{numerics})).}
 \label{geometry}
\end{figure}
\begin{figure}
\protect \centerline{\epsfxsize=4.5in \epsfbox {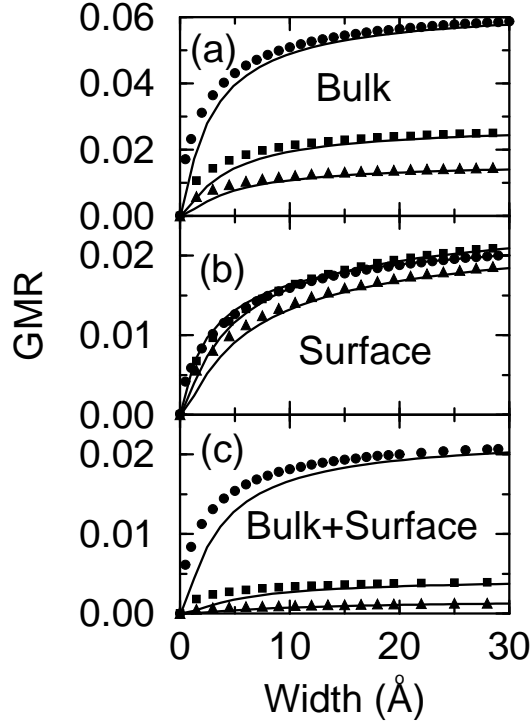}}
\vskip .6cm
\protect\caption{
 Giant magnetoresistance of a three wire structure as a function of width
 for (a) bulk scattering, (b)
 surface scattering and (c) both bulk and surface scattering. 
 The symbols refer to different film thicknesses,
 $t_{\rm PM} = t_{\rm FM} = $
 8\AA \,(circle),
 20\AA \,(box), and 
 30\AA \,(triangle).
 In all cases the
 GMR decreases as we laterally confine the multilayers. 
 The origin of this decrease in the GMR can be understood in 
 terms of changing the effective mean free paths in the wires.
 As the wire width is reduced the effective mean free path
 within each wire decreases.
 To make this more quantitative we obtain an effective
 mean free path for each wire and use these mean free
 paths in a multilayer calculation (infinite width wire).
 The results, which are shown as the solid lines, are in
 good agreement with the exact calculation (symbols).
 For cases (a) and (c) 
 the mean free paths are chosen for a Co/Cu/Co structure,\cite{gurney}
 which has $L^\uparrow$=55\AA, $L^\downarrow$=10\AA, and $L^{\rm PM}$=226\AA.
 For case (b), in which only surface scattering contributes to the
 GMR, we use
 $L^\uparrow = L^\downarrow$=17\AA \, and $L^{\rm PM}$=226\AA.
 For cases (b) and (c)
 the transmission coefficients at the interfaces are 
 taken to be $T^\uparrow$=0.8, $T^\downarrow$=0.4,\cite{stiles} 
 while for case (a), where the GMR is due only to bulk scattering,
 we take $T^\uparrow$=$T^\downarrow$=0.6.
 The thicknesses 
 for wire 2 are chosen to be that of Cu when the Co slabs are 
 antiferromagnetically coupled,\cite{parkin} and for simplicity we choose
 the Co layers to have the same thickness as the Cu. 
 In all cases
 the sides and the interfaces have a specularity factor $S$=0.9.}
 \label{gmr-width}
 \end{figure}
 
\begin{figure}
\protect \centerline{\epsfxsize=3.5in \epsfbox {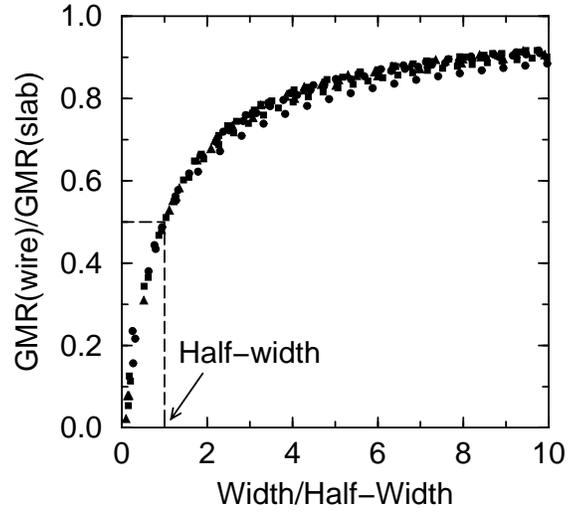}}
\vskip .6cm
\protect \caption{
 Rescaled giant magnetoresistance as a function of width.
 For large width the GMR for a wire approaches that for
 an ordinary unconfined multilayer, GMR(slab).
 We define
 the width at which the GMR is reduced to half of GMR(slab)
 as the half-width.
 Rescaling the giant magnetoresistance
 by GMR(slab) and the width by the
 half-width, the GMR vs. width curves of Fig. 2 
 fall onto a single curve.  The symbols are the same ones used
 in Figs. 2(a)-(c).}
\label{scaling}
\end{figure}
\newpage
 \begin{figure}
\protect \centerline{\epsfxsize=3.5in \epsfbox {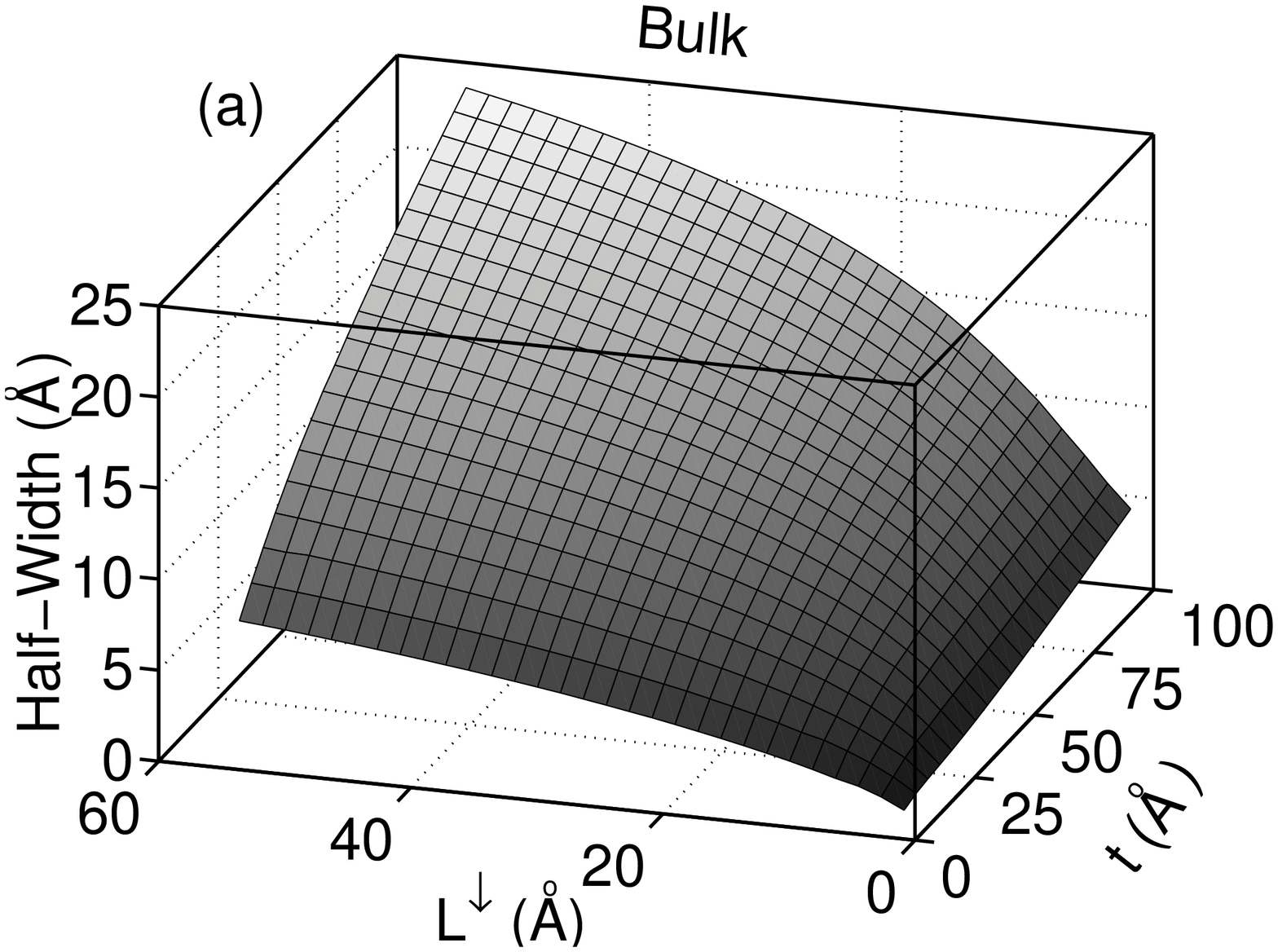}}
\protect \centerline{\epsfxsize=3.5in \epsfbox {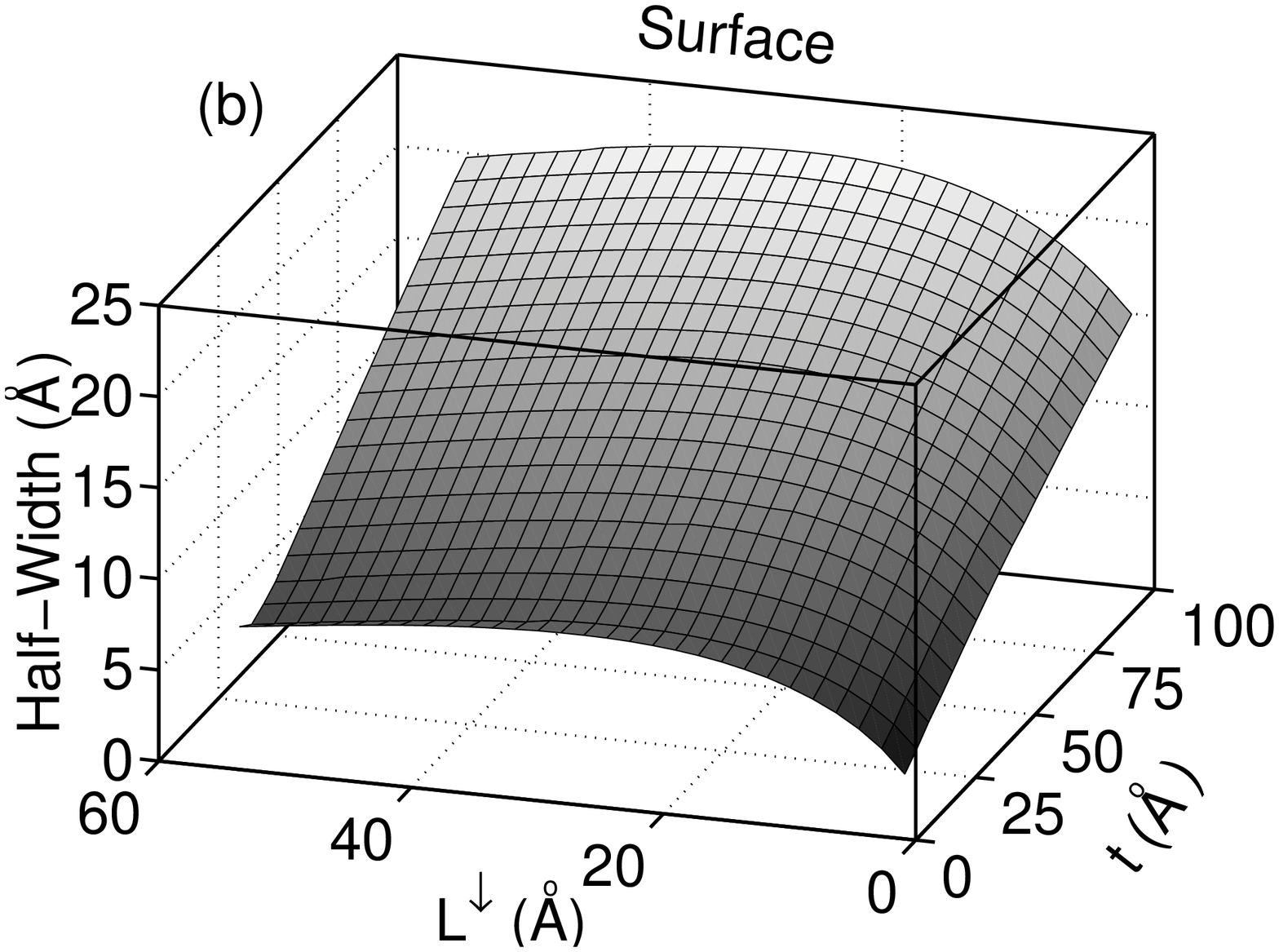}}
\protect \centerline{\epsfxsize=3.5in \epsfbox {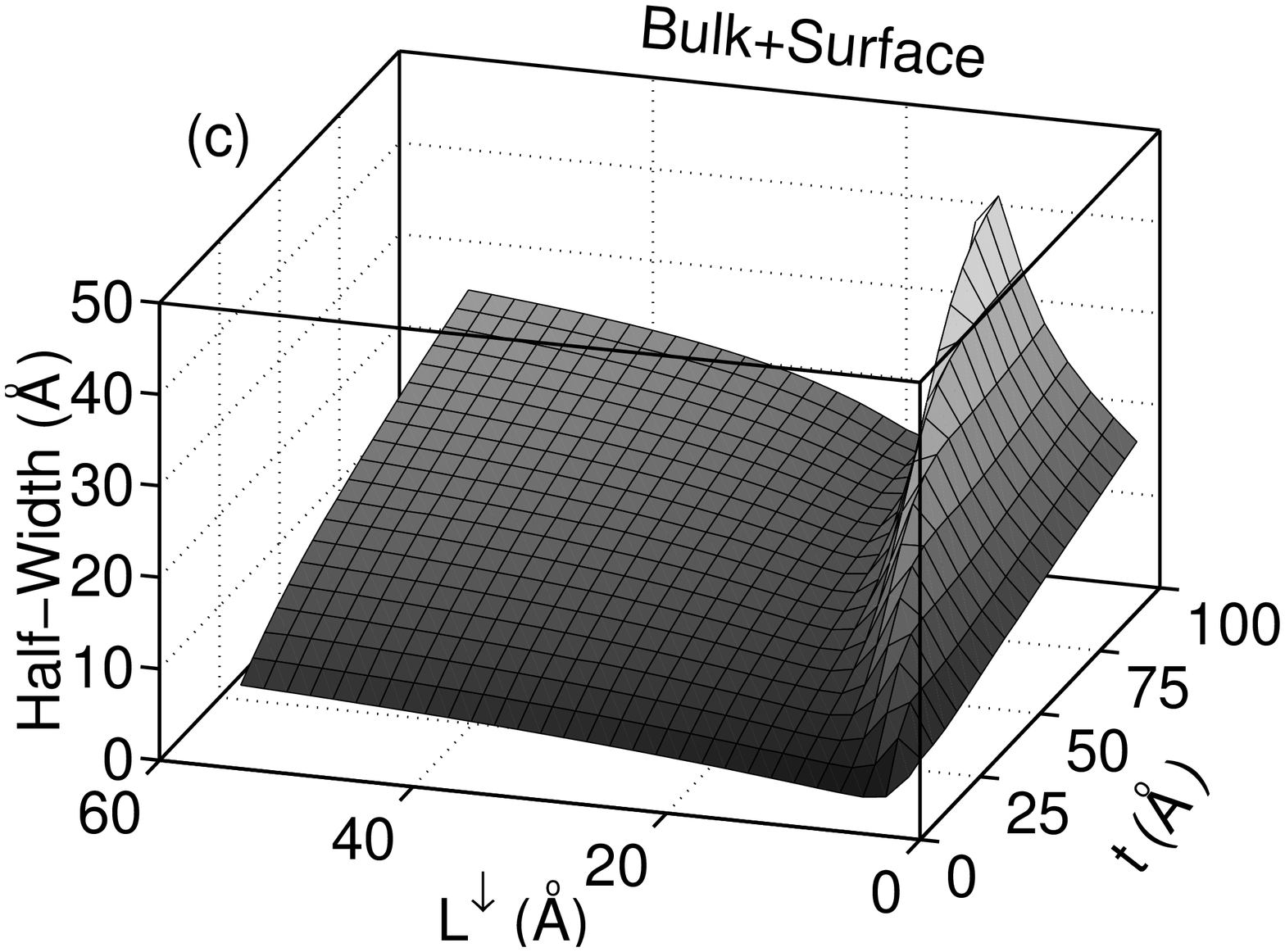}}
\vskip .6cm
\protect
\caption{ Half-width as a function of the thickness and mean free
 path in the ferromagnet ($L^\downarrow$) for (a) bulk scattering, (b) surface
 scattering, and (c) bulk and surface scattering. 
 In Fig. 4(a) and (c) the ferromagnetic spin-up and spin-down mean free paths
 are kept at a fixed ratio,
 $L^{\uparrow} / L^{\downarrow}=5.5$,
 which is the same ratio used in Figs. 2(a) and (c).
 In the surface scattering case of Fig. 4(b) $L^{\uparrow}$ equals
 $L^{\downarrow}$.
 The half-width depends both on 
 the thickness and the mean free path
 for all three cases. 
 In case (a), bulk scattering,
 the half-width increases roughly linearly with
 the thickness and the mean free path,
 while in case (b), surface scattering,
 the half-width depends primarily on the thickness of the layers.
 The general case when both bulk and surface scattering are 
 important (case (c)) can lead to complex dependence on the
 mean free path and film thickness.  The feature at small
 $L^{\downarrow}$ in (c) is associated with the fact that
 the GMR has a local minimum in this region (see discussion in text).
 Except for $t$ and $L^{\downarrow}$, which vary,
 the parameters used in (a), (b), and (c) are the same
 as in Fig. 2.}
\label{half-width}
\end{figure}
\newpage
\begin{figure}
\protect \centerline{\epsfxsize=3.5in \epsfbox {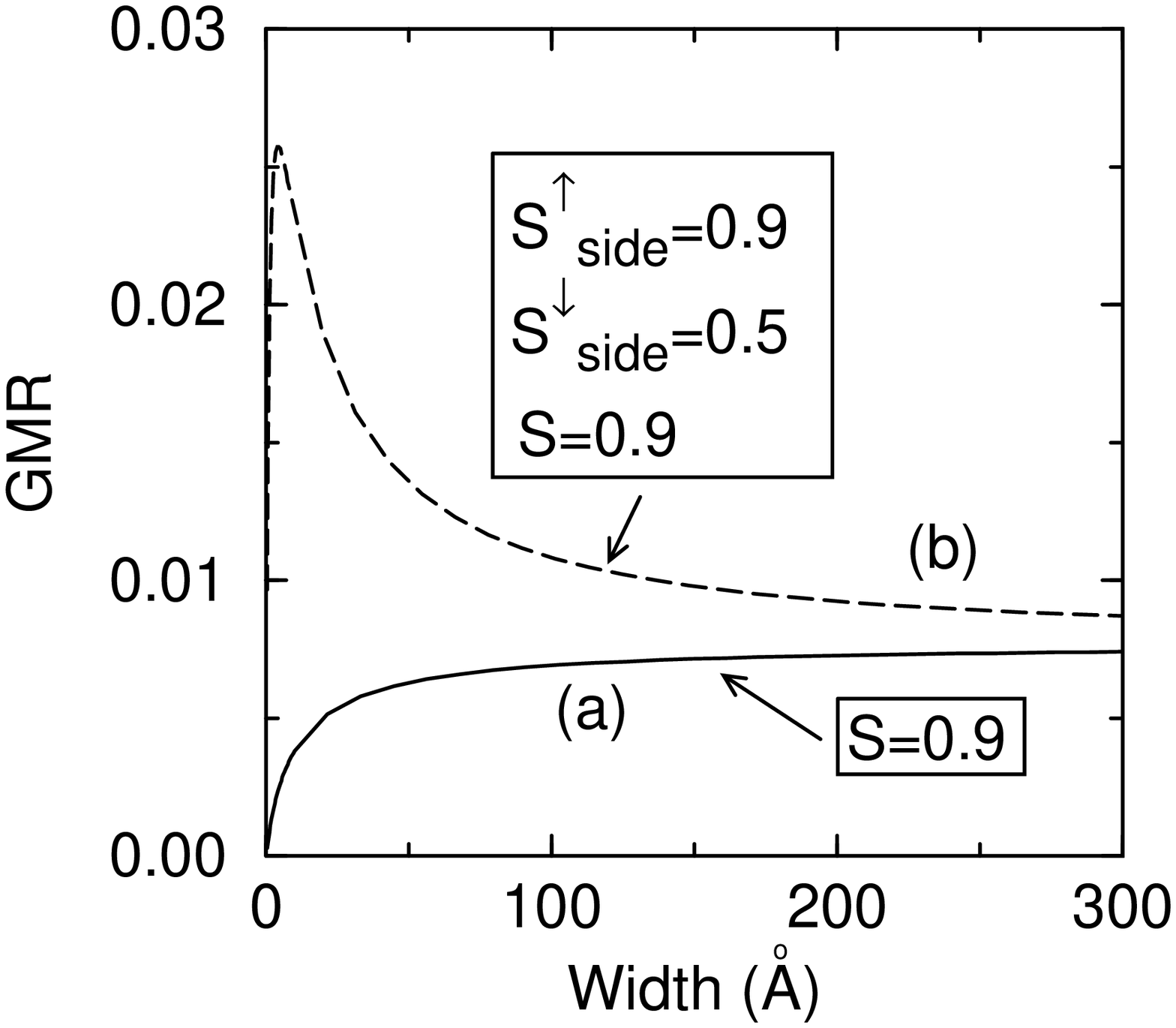}}
\vskip .6cm
\protect \caption{Giant magnetoresistance as a function of width for
 (a) spin-independent scattering at the sides and interfaces, 
 (b) spin-dependent scattering at the sides of the ferromagnet.
 For (a), the GMR decreases with decreasing width (solid line), 
  whereas for (b), the GMR
  increases with decreasing width (dashed line). 
  The increase of the GMR with reduced
  width is due to the spin-down mean free path decreasing
  faster than the spin-up one. For (a), we choose the same parameters
  as of Fig. 2, except the mean free paths for the spin-up and down electrons 
  in the ferromagnet are
  200\AA \, and 100\AA , respectively.
  For (b), we use the same parameters except that the specularity factors
  for the sides of the 
  ferromagnet are spin-dependent: 
  $S^{\uparrow}_{\rm side}$=0.9 and $S^{\downarrow}_{\rm side}$=0.5.  
  Note that the GMR eventually goes to zero for small enough widths
  because the effective mean free path in the paramagnet becomes much smaller
  than the thickness of the paramagnetic layer.}
 \label{special}
 \end{figure}


\begin{thebibliography}{19}
\bibitem{ptoday} Phys. Today, {\bf 48}, 24 (1995).
\bibitem{baibich} M. N. Baibich, J. M. Broto, A. Fert, F. Nguyen Van Dau,
F. Petroff, P. Etienne, G. Creuzet, A. Friederich, and J. Chazelas, Phys.
Rev. Lett. {\bf 61}, 2472 (1988); A. Barth\'{e}l\'{e}my, A. Fert, M. N. Baibich,
S. Hadjoudj, F. Petroff, P. Etienne, R. Cabanel, S. Lequien, and G. Creuzet,
J. Appl. Phys. {\bf 67}, 5908 (1990). 
\bibitem{grunberg} G. Binach, P. Grunberg, F. Saurenbach, and W. Zinn,
Phys. Rev. B {\bf 39}, 4828 (1989).
\bibitem{white} R. L. White, IEEE Trans. Magn. {\bf 30}, 346 (1994).
\bibitem{fuchs} K. Fuchs Proc. Camb. Philos. Soc. {\bf 34}, 100 (1938).
\bibitem{sond} E. H. Sondheimer, Adv. Phys. {\bf 1}, 1 (1952).
\bibitem{levy} P. M. Levy: ``Giant Magnetoresistance in Magnetic Layered
and Granular Materials'', in {\it Solid State Physics}, ed. by F. Seitz,
D. Turnbull, H. Ehrnereich, Vol {\bf 47}, p. 367 (Academic Press, New York, 
1994).
\bibitem{levy2} H. E. Camblong and P. M. Levy, Phys. Rev. Lett. {\bf 69},
2835 (1992); J. Appl. Phys. {\bf 73}, 5333 (1993); H. E. Camblong,
 Phys. Rev. B {\bf 51}, 1855 (1995).
\bibitem{fert} A. Fert and P. Bruno: ``Interlayer Coupling and 
Magnetoresistance in Multilayers'', in {\it Ultrathin Magnetic Structures II},
ed. by B. Heinrich, J. A. C. Bland, p. 82 (Springer-Verlag, Berlin Heidelberg,
1994); S. S. P. Parkin: ``Giant Magnetoresistance and Oscillatory 
Interlayer Coupling in Polycrystalline Transition Metal Multilayers'', 
{\it ibid}, p. 148;
A. Fert: ``Transport Properties of Thin Metallic Films
and Multilayers'', in {\it Science and Technology of Nanostructured
Magnetic Materials}, ed. by G. C. Hadjipanayis, G. A. Prinz (Plenum Press,
London, 1991) p. 221.
\bibitem{visscher} P. B. Visscher, Phys. Rev. B, {\bf 49}, 3907 (1994).
\bibitem{zhang} X.-G. Zhang and W. H. Butler, Phys. Rev. B, {\bf 51}, 
10085 (1995). 
\bibitem{c-b} R. E. Camley and J. Barna\'{s}, Phys. Rev. Lett. {\bf 63}, 
664 (1989); J. Barna\'{s}, A. Fuss, R. E. Camley, P. Grunberg, and W. Zinn,
 Phys. Rev. B, {\bf 42}, 8110 (1990).
\bibitem{childress} Y. D. Park, J. A. Caballero, A. Cabbibo, J. R. Childress,
 H. D. Hudspeth, T. J. Schultz, and F. Sharifi, J. Appl. Phys. {\bf 81}, 
 4717 (1997).
\bibitem{chambers} R. G. Chambers, Proc. Roy. Soc. A, {\bf 202}, 378 
(1950).
\bibitem{lenk} R. Lenk and A. Kn\'{a}bchen, J. Phys.: Condens. Matter 
{\bf 5}, 6563 (1993).
\bibitem{hood} R. Q. Hood and L. M. Falicov, Phys. Rev. B, {\bf 46},
 8287 (1992).
\bibitem{stiles} M. D. Stiles, J. Appl. Phys. {\bf 79}, 5805 (1996).
\bibitem{gurney} Bruce A. Gurney, Virgil S. Speriosu, Jean-Pierre
 Nozieres, Harry Lefakis, Dennis R. Wilhoit, and Omar U. Need, Phys. Rev.
  Lett. {\bf 71}, 4023 (1993).
\bibitem{parkin} S. S. P. Parkin, R. Bhadra, and K. P. Roche, Phys. Rev.
Lett. {\bf 66}, 2152 (1991). 
\end{thebibliography}
\end{document}